\documentclass[twocolumn,showpacs,pra,amsmath,amssymb]{revtex4}
\usepackage{graphicx}% Include figure files
\usepackage{dcolumn}% Align table columns on decimal point
\usepackage{bm}% bold math

\begin{document}

\title{ On-demand single-photon state generation via nonlinear absorption }
\author{Tao \surname{Hong}}
\affiliation{NTT Basic Research Laboratories, NTT Corporation, 3-1, Morinosato-Wakamiya, Atsugi-shi, Kanagawa 243-0198, Japan}
\affiliation{Department of Physics, University of Washington, Seattle, Washington 98195.}
\author{Michael W. \surname{Jack}}
\affiliation{NTT Basic Research Laboratories, NTT Corporation, 3-1, Morinosato-Wakamiya, Atsugi-shi, Kanagawa 243-0198, Japan}
\affiliation{Department of Physics and Astronomy and Rice Quantum Institute, Rice University, Houston, Texas 77251}
\author{Makoto \surname{Yamashita}}
\affiliation{NTT Basic Research Laboratories, NTT Corporation, 3-1, Morinosato-Wakamiya, Atsugi-shi, Kanagawa 243-0198, Japan}
%\date{\today}

\begin{abstract}
We propose a method for producing on-demand single-photon states based on collision-induced exchanges of photons and unbalanced linear absorption between two single-mode light fields. These two effects result in an effective nonlinear absorption of photons in one of the modes, which can lead to single photon states. A quantum nonlinear attenuator based on such a mechanism can absorb photons in a normal input light pulse and terminate the absorption at a single-photon state. Because the output light pulses containing single photons preserve the properties of the input pulses, we expect this method to be a means for building a highly controllable single photon source.
\end{abstract}

\maketitle 

Generation and control of single-photon states are not only of fundamental interest\cite{Fundamental}, but are also a crucial component in realizing many quantum information devices\cite{Communication}. Secure key distribution in quantum cryptography requires an optical source that can emit a train of pulses each of which, to a very high probability, contains only one photon\cite{Communication}.  Photons are said to exhibit antibunching if the number fluctuations in the pulses are reduced below the classical Poissonian distribution\cite{Fundamental}. In addition, it has been shown that one can implement quantum computation using only linear optical elements and photodetectors, given a suitable single-photon source\cite{LOCC}. In this case, however, one requires not only one photon per pulse but also the more stringent condition that consecutive photons have identical wave packets so that they act as indistinguishable photons in a multi-photon interference experiment\cite{Indistinguishable}.

Recently, several single-photon sources based on single quantum systems have been demonstrated using different devices, including dye molecules\cite{Molecules}, quantum wells\cite{Wells}, color centers\cite{ColourCentres}, trapped atoms\cite{Atoms} and quantum dots \cite{Dots,Santori}. In particular, quantum dots and color centers have already been successfully applied to quantum cryptography \cite{Application}. At present, only quantum dots have exhibited the ability to emit photons that are largely identical\cite{Indistinguishable}, as required for quantum computation \cite{LOCC}. However, the utility of 
these quantum dots as single-photon sources are limited by decoherence 
due to phonon-scattering which causes the spontaneously-emitted photons 
to become distinguishable\cite{Indistinguishable,Phononscattering}.
On the other hand, lasers, combined with well-developed light modulation techniques, have many advantages in producing identical pulses at the Fourier transform limit.
Ideally, we would like to be able to attenuate the laser intensity until only one photon is left in each pulse. So far, no mechanisim based on absorption has been found useful in producing on-demand single photons, i.e. generating single photons at arbitrary predetermined times within short time intervals.  Ordinary linear absorption cannot produce on-demand single photons because the photon number fluctuations of even very faint laser pulses passing through a linear attenuator still obeys Poissonian statistics \cite{Fundamental}.  Detection of one of the twin photons created by parametric down-conversion is a possible method of  producing single photons, but the generation time of a photon pair is randomly distributed and therefore does not satisfy the on-demand condition \cite{DownConversion}. Two-photon absorption can also attenuate a multi-photon state to a single-photon state, but the probability of producing a single-photon state is at most $50\%$ \cite{TwoPhotonAbsorption}. 

In this paper, we propose a new method for generating on-demand single-photon states by nonlinear absorption. Since there is no difficulty in obtaining well-controlled light pulses by ordinary laser techniques, we simply assume that ``on demand'' one can send a laser pulse, which contains many photons, through a quantum third-order nonlinear attenuator. In the attenuator, the Hamiltonian for mode $\hat{a}$ of the light field contains an interaction term of the form, 
$\hbar U\hat{a}^{\dag}\hat{a}\hat{a}\hat{G}^{\dag}+adj.$, where $\hat{G}$ denotes the annihilation operator of the vacuum bath field, and $U$ denotes the coupling coefficient. Intuitively, this damping process leads to a cascade down the number states of mode $\hat{a}$ terminating in one photon: $\hat{a}^{\dag}\hat{a}\hat{a}|2\rangle \rightarrow |1\rangle$. This should be contrasted with two-photon absorption where the even number states terminate in the vaccum $\hat{a}^{2}|2\rangle \rightarrow |0\rangle$, such that the final state is a equal mixture of $1$ and $0$ photons. In addition, we analyze the degredation of  this mechanism in the presence of other loss mechanisms to clarify the requirement for realizing the quantum nonlinear attenuator experimentally. 

The basic idea of the proposal is based on nonlinear coupling and extremely unbalanced linear absorption of two-mode light fields. We consider two quantized light fields in the attenuator as two independent normalized modes, A and B (if they are degenerate in frequency, then they are orthogonal). These two modes couple to four independent vacuum fields Q, R, S and T. The total Hamiltonian of the system can be written as
\begin{eqnarray}
\hat{H}= \hat{H}_{0}+\hat{H}_{I}+\hat{H}_{i}+\hat{H}_{R}
\label{eq1}
\end{eqnarray}
where 
$\hat{H}_{0}=\hbar\omega_{a}\hat{a}^{\dag}\hat{a}
+\hbar\omega_{b}\hat{b}^{\dag}\hat{b}$, 
$\hat{H}_{R}=\sum_{i}\hbar\omega_{qi}\hat{q}_{i}^{\dag}\hat{q}_{i}
+\sum_{i}\hbar\omega_{ri}\hat{r}_{i}^{\dag}\hat{r}_{i}
+\sum_{i}\hbar\omega_{si}\hat{s}_{i}^{\dag}\hat{s}_{i}
+\sum_{i}\hbar\omega_{ti}\hat{t}_{i}^{\dag}\hat{t}_{i}$,
$\hat{H}_{i}=\hat{a}^{\dag}\sum_{i}\hbar f_{i}\hat{q}_{i}
+\hat{b}^{\dag}\sum_{i}\hbar g_{i}\hat{r}_{i}
+\hat{a}^{\dag}\hat{a}^{\dag}\sum_{i}\hbar u_{i}\hat{s}_{i}
+\hat{a}^{\dag}\hat{a}^{\dag}\hat{a}^{\dag}\sum_{i}\hbar w_{i}\hat{t}_{i}
+ adj.$
and a nonlinear interaction term
\begin{eqnarray}
\hat{H}_{I}=\hbar U_{1}\hat{a}^{\dag}\hat{a}^{\dag}\hat{a}\hat{a}
+\hbar U_{2}\hat{b}^{\dag}\hat{b}^{\dag}\hat{b}\hat{b}
+\hbar U_{3}\hat{a}^{\dag}\hat{a}\hat{b}^{\dag}\hat{b} \\ \nonumber
+[\hbar U_{4}\hat{a}^{\dag}\hat{a}\hat{a}\hat{b}^{\dag} 
+\hbar U_{5}\hat{a}\hat{b}^{\dag}\hat{b}^{\dag}\hat{b} 
+\hbar U_{6}\hat{a}\hat{a}\hat{b}^{\dag}\hat{b}^{\dag} 
+ adj.].
\label{eq2}
\end{eqnarray}
Here $\hat{a}$ and $\hat{b}$ are Boson annihilation operators of the modes A and B, and $\hat{q }_i $, $\hat{r }_i $, $\hat{s}_i$  and $\hat{t}_i$ are Boson annihilation operators of the $i$-th modes of the reservoirs Q, R, S and T, respectively.  All these operators satisfy the bosonic commutation relations. $\omega_a$, $\omega_b$, $\omega_{qi}$, $\omega_{ri}$, $\omega_{si}$ and $\omega_{ti}$ are the eigen-frequencies of the above mentioned modes, as denoted by their suffixes, and $f_{i}=f(\omega_{qi})$, $g_{i}=g(\omega_{ri})$, $u_{i}=u(\omega_{si})$ and $w_{i}=w(\omega_{ti})$ are coupling coefficients of the modes A and B with the $i$-th modes of the four vacuum fields, respectively. $U_{1}\sim U_{6}$ denote the frequencies of interaction strengths between photons in the modes A and B.  The values of $U_{1}\sim U_{3}$ are real due to the Hermitian nature of the Hamiltonian.

The essential term for our nonlinear attenuator scheme is 
$U_{4}\hat{a}^{\dagger}\hat{a}\hat{a}\hat{b}^{\dagger}$, where $U_{4}$ 
is assumed to be nonvanishing. In terms of the spatial mode functions this coefficient is given by $U_{4}\propto\int d^{3}{\bf x}|Q_{a}({\bf x})|^{2}Q_{a}({\bf 
x})Q_{b}^{*}({\bf x})$, where $Q_{a}({\bf x})$ and $Q_{b}({\bf x})$ are the spatial mode functions corresponding to the operators $\hat{a}$ and $\hat{b}$. This term is often neglected in conventional 
nonlinear optics because if $Q_{a}({\bf x})$ and $Q_{b}({\bf x})$ are 
plane waves with the same frequency but different momentum, then this term does no satisfy 
momentum conservation. However, if there is some momentum transfer 
from photons to the medium, then $U_{4}$ has a finite value. For 
example, if $Q_{a}({\bf x})$ or $Q_{b}({\bf x})$ correspond to localized 
modes or if there is a momentum recoil of the atoms in the gaseous 
nonlinear media that mediates the interaction between photons in the two modes (In this case, the wave function of the atoms must be added into the integral of $U_{4}$.).
Alternatively, if the modes A and B differ in frequency, we assume that the two mode functions $Q_{a}({\bf x})$ and $Q_{b}({\bf x})$ overlap significantly so that $U_{4}$ is sufficiently large and there is still a non-vanishing average photon transfer between the two modes.

Next, we reduce the above total Hamiltonian and focus our discussion on the quantum dynamical behavior of the mode A. From Eq. (\ref{eq1}), we can derive the Langevin equation for the annihilation operator $\hat{b}$.  Assuming the coupling coefficient, $g_{i}$, between the mode B and the vacuum field R is much larger than other coupling coefficients as well as photon interacting strengths $U_{2}\sim U_{6}$, we
can neglect all other coupling terms in the Langevin equation and use the adiabatic approximation to derive $\hat{b}(t)\approx 2\hat{G}(t)/\Gamma_{b}$, where $\hat{G}(t)=-i\sum_{j}{g_{j}\hat{r}_{j}(t_{0})e^{-i\omega_{rj}(t-t_{0})}}$, $\Gamma_{b}=2\pi D_{r}(\omega_{b})|g(\omega_{b})|^{2}$ and $D_{r}(\omega)$ denotes the mode density of the vacuum field R at frequency $\omega$. It is evident that under the perturbation of the vacuum field the actual frequency spectrum of the mode B is much broadened. Then from the total Hamiltonian, we can get an effective Hamiltonian for the mode A,
\begin{equation}
\hat{H}_{\rm eff}=\hat{H}_{0e}+\hat{H}_{ie}+\hat{H}_{R},
\label{eq3}
\end{equation}
where $\hat{H}_{0e}=\hbar\omega_{a} \hat{a}^{\dag}\hat{a}+\hbar U_{1}\hat{a}^{\dag}\hat{a}^{\dag}\hat{a}\hat{a}$ and 
$\hat{H}_{ie}=\hbar U_{4}\hat{a}^{\dag}\hat{a}\hat{a}{2\hat{G}^{\dag}}/{\Gamma_{b}}   
+\hat{a}^{\dag}\sum_{i}\hbar f_{i}\hat{q}_{i}
+\hat{a}^{\dag}\hat{a}^{\dag}\sum_{i}\hbar u_{i}\hat{s}_{i}
+\hat{a}^{\dag}\hat{a}^{\dag}\hat{a}^{\dag}\sum_{i}\hbar w_{i}\hat{t}_{i} + adj.$. 
Note that in the derivation we keep only the first-order terms in  $1/\Gamma_{b}$, because the damping rate $\Gamma_{b}$ is very large and the photon number of the mode A is very small. From this effective Hamiltonian and making the Markov approximation, we can derive a master equation for the mode A:
\begin{eqnarray}
\frac{d\rho(t)}{dt}=
-\frac{\Gamma_{e}}{2}
[\hat{a}^{\dag}\hat{a}^{\dag}\hat{a}\hat{a}^{\dag}\hat{a}\hat{a}\rho(t) -\hat{a}^{\dag}\hat{a}\hat{a}\rho(t)\hat{a}^{\dag}\hat{a}^{\dag}\hat{a}]
\nonumber \\ 
-\frac{\Gamma_{q}}{2}
[\hat{a}^{\dag}\hat{a}\rho(t)-\hat{a}\rho(t)\hat{a}^{\dag}]  
-\frac{\Gamma_{s}}{2}
[\hat{a}^{\dag}\hat{a}^{\dag}\hat{a}\hat{a}\rho(t)-\hat{a}\hat{a}\rho(t)\hat{a}^{\dag}\hat{a}^{\dag}]
\nonumber \\
-\frac{\Gamma_{t}}{2}
[\hat{a}^{\dag}\hat{a}^{\dag}\hat{a}^{\dag}\hat{a}\hat{a}\hat{a}\rho(t)-\hat{a}\hat{a}\hat{a}\rho(t)\hat{a}^{\dag}\hat{a}^{\dag}\hat{a}^{\dag}]
+adj.
\label{eq4}
\end{eqnarray}
where $\Gamma_{q}=2\pi D_{q}(\omega_{a})|f(\omega_{a})|^{2}$,
$\Gamma_{s}=2\pi D_{s}(\omega_{a})|u(\omega_{a})|^{2}$,
$\Gamma_{t}=2\pi D_{t}(\omega_{a})|w(\omega_{a})|^{2}$ and
$\Gamma_{e}=\Gamma_{a}|{2U_{4}}/{\Gamma_{b}}|^{2}$ are
the linear damping rate, the two-photon damping rate, the three-photon damping rate and the effective nonlinear damping rate, respectively. Note that the mode density $D_{r}$ (or $D_{q}$, or $D_{s}$, or $D_{t}$) and mode coupling strength $g$ (or $f$, or $u$, or $w$) are assumed to be broad and flat around the frequency $\omega_a$, so that the variation of the product due to the frequency shift produced by the nonlinear term $\hbar U_{1}\hat{a}^{\dag}\hat{a}^{\dag}\hat{a}\hat{a}$ 
is negligible. The relation of $\Gamma_{e}$ with the linear damping rate
$\Gamma_{a}=2\pi D_{r}(\omega_{a})|g(\omega_{a})|^{2}$ and $U_4$ indicates that the effective nonlinear damping of photons in the mode A comes from the combination of photon-interaction-induced transfer from the mode A to the mode B and the strong linear damping of photons in the mode B. 

In an ideal attenuator, where the damping rates $\Gamma_{q}=\Gamma_{s}=\Gamma_{t}=0$, the final steady state of the light field can be a single-photon state. From Eq. (\ref{eq4}), we can derive the equation for the matrix elements of $\rho (t)=\sum_{k, l}\rho_{kl}|k\rangle\langle l|$ as follows, 
\begin{eqnarray}
\frac{d\rho_{kl}}{dt}=-(\Gamma_{e}/2)[k(k-1)^{2}+l(l-1)^{2}]\rho_{kl}\nonumber \\
+\Gamma_{e}\sqrt{k^{2}(k+1)l^{2}(l+1)}\rho_{k+1,l+1}
\label{eq5}
\end{eqnarray}
>From this equation, we see that the density matrix elements with $k$ or $l >1$ always damp, and the damping of the density matrix elements at larger photon numbers leads to growth of the density matrix elements at smaller photon numbers. If both $k$ and $l$ equal $1$ or $0$, then the damping rates disappear.  $\rho_{11}$ still grows due to the damping of $\rho_{22}$, but $\rho_{00}$ decouples from other density matrix elements and its value is determined only by the initial states of the light field. Thus so long as the initial state has a finite number of photons on average and has zero probability at the zero photon number state, then after a sufficient time of evolution in the attenuator, the final steady state of the light field must be a single-photon state. 

\begin{figure}[h]
\includegraphics[width=6cm, height=8cm, angle=270]{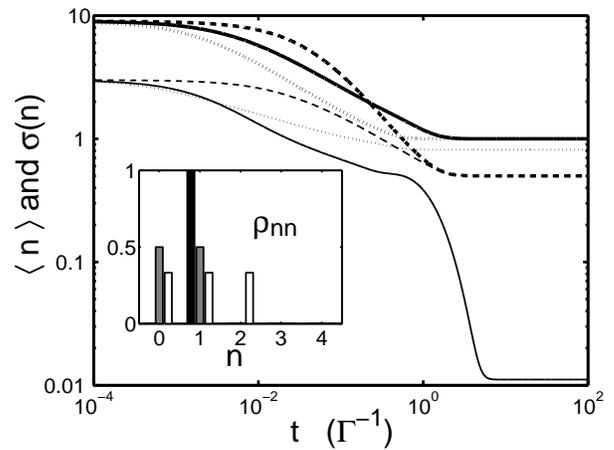} 
\caption{Dependence of average photon number $\langle n\rangle$ (thick lines) and standard deviation $\sigma(n)$ (thin lines) on time $t$ for three damping processes: 
(1) $\Gamma=\Gamma_{e}$, $\Gamma_{q}=\Gamma_{s}=\Gamma_{t}=0\Gamma$ (solid lines); 
(2) $\Gamma=\Gamma_{s}$, $\Gamma_{e}=\Gamma_{q}=\Gamma_{t}=0\Gamma$ (dashed lines); 
(3) $\Gamma=\Gamma_{t}$, $\Gamma_{e}=\Gamma_{q}=\Gamma_{s}=0\Gamma$ (dotted lines); 
Inset shows the probability distribution of the photon number $n$ at $t=100\Gamma^{-1}$: black bars for process (1), gray bars for process (2), white bars for process (3). The three processes start from the same coherent state $|\alpha\rangle$, where $\alpha=3$.} 
\label{fig1}
\end{figure}

The dynamics of the nonlinear attenuation is more clearly demonstrated by numerical solution of the master equation (\ref{eq4}). 
Solid lines in Fig. \ref{fig1} show the time evolution of the average photon number $\langle n\rangle=\langle \hat{a}^{\dag}\hat{a}\rangle$ and standard deviation $\sigma(n)=\langle(\hat{a}^{\dag}\hat{a})^{2}-\langle \hat{a}^{\dag}\hat{a}\rangle^{2}\rangle^{1/2}$ in a pure effective nonlinear absorption process starting from a coherent state $|\alpha\rangle$.
We clearly observe that the average photon number $\langle n\rangle$ asymptotically approaches $1$ and the standard deviation $\sigma(n)$ is simultaneously greatly reduced when the evolution time is sufficient long in comparison with $1/\Gamma_{e}$. 
Because the standard deviation $\sigma(n)$ is much smaller than the average photon number $\langle n\rangle$, the final state is approximately squeezed to a Fock state, in which the photon number $n=1$. The probability for achieving this single-photon state at $t=100\Gamma^{-1}$, as shown by the black bars in the inset, demonstrating this point further.
Thus such a nonlinear attenuator can attenuate the light field nonlinearly such that the attenuation terminates at one photon. 
This is just the property that is required for an on-demand single-photon source. In contrast, a process based on two-photon absorption (or three-photon) absorption cannot produce single photons on demand. Dashed lines (or dotted lines) in Fig. \ref{fig1} show that the average photon number is less than $1$ and the standard deviation remains very large even at $t\gg \Gamma^{-1}$ in a pure two-photon (or three-photon) absorption process. In the two-photon (or three-photon) absorption process, the probability for achieving single photons is actually only $1/2$ (or $1/3$) at most, as shown by the gray bars (or the white bars). In this situation, the standard deviation of photon numbers in the effective nonlinear absorption process does no reach zero, because of the initial occupation of the zero photon state. To achieve single photons with high reliability, a requirement for the initial coherent state with sufficiently large average photon number is therefore necessary. 

\begin{figure}[h]
\includegraphics[width=6cm, height=8cm, angle=270]{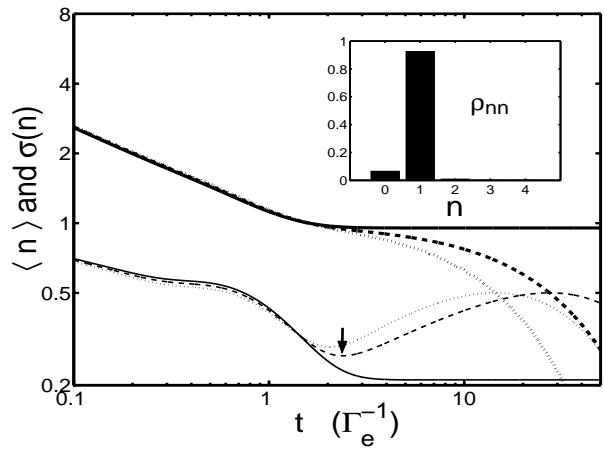} 
\caption{Dependence of average photon numbers $\langle n\rangle$ (thick lines) and standard deviations $\sigma(n)$ (thin lines) on time $t$ for three damping processes: 
(1) $\Gamma_{q}=0.05\Gamma_{e}$ and $\Gamma_{s}=\Gamma_{t}=0\Gamma_{e}$ (solid lines); 
(2) $\Gamma_{q}=\Gamma_{t}=0\Gamma_{e}$ and $\Gamma_{s}=0.05\Gamma_{e}$ (dotted lines); 
(3) $\Gamma_{q}=\Gamma_{s}=0.025\Gamma_{e}$ and $\Gamma_{t}=0\Gamma_{e}$ (dashed lines). Inset shows the probability distribution of the photon number $n$ at the local minimum of $\sigma(n)$ in process (3) at $t\approx 2.5 \Gamma_{e}^{-1}$, as denoted by the arrow. The three processes start from the same coherent state $|\alpha\rangle$, where $\alpha=3$.} 
\label{fig2}
\end{figure}

In a real system, linear absorption and two-photon absorption are not likely to be eliminated completely, thus it is necessary to consider their negative influence on the single photon generation process. Figure \ref{fig2} shows the damping processes in cases of mixed absorptions. When the damping time $t\gg \Gamma_{e}^{-1}$, two-photon absorption decreases the average photon number, which asymptotically approaches a constant less than $1$, and increases the standard deviation $\sigma(n)$, as shown by solid lines. Hence the probability of single-photon state is decreased due to the presence of two-photon absorption.  Linear absorption appears to produce much stronger negative influence on the single photon generation process. It decreases the average photon number to zero and increases the fluctuation of the photon number more significantly if the damping time $t\gg \Gamma_{e}^{-1}$, as shown by the dotted lines. However, the local minimum of $\sigma(n)$ indicates that it is still possible to achieve single-photon states with fairly large probability at an intermediate time $t$ slightly larger than $\Gamma_{e}^{-1}$. In particular, the inset shows that even when linear absorption and two-photon absorption 
coexists with the effective nonlinear absorption, it is still possible to
achieve single-photon states with very large probability at intermediate times.

Next, we summarize the medium requirements and give some suggestions for experimental realization of the above theoretical model.  First, because the effective nonlinear damping rate $\Gamma_{e}$ is proportional to the photon interaction strength $|U_{4}|^{2}$, the medium must provide strong Kerr nonlinearity so that the effective nonlinear damping rate $\Gamma_{e}$ overcomes other damping rates; second, the medium must provide very different linear absorption between the mode A and the mode B, and in particular, the linear absorption for the mode A must be much smaller than the effective nonlinear absorption. One can realize different linear absorption of the two modes by using narrow band optical elements (like Fabry-Perot cavities), spatial filters, or polarizing elements (like optical isolators) to transmit the mode A and attenuate the mode B in the case where the mode A and the mode B differ in frequency, are orthogonal in spatial distribution, or are orthogonal in polarization, respectively. Although one type of medium is not usually adequate for making any optical structure (such as a waveguide) to confine localized modes A and B, we think electro-magnetically induced transparency atomic vapor may be a good candidate for at least one of the composition mediums because, in the case of a four-level medium, it has been found to provide a giant optical Kerr effect and simultaneously vanishing linear and two-photon absorption \cite{EITMedium}. In addition, because the medium is a vapor, momentum transfer from photons to atoms can occur which can mediate an interaction between the mode A and the mode B in the case when they are unlocalized modes of different momenta. If the atoms have more magnetic sublevels, the vapor can also mediate the interaction between optical modes of different polarization so that the mode A and the mode B can be orthogonal in polarization.

In conclusion, we have proposed a new method for producing on-demand single-photon states. The method is based on collision-induced exchanges of photons and unbalanced linear absorption between two-mode light fields. We have shown that the two effects result in an effective nonlinear absorption of photons in one of the modes, which can lead to single photon states. Thus a quantum nonlinear attenuator based on such a mechanism can absorb photons in a normally controlled input light pulse and terminate the absorption at a single-photon state which inherits many properties of the input pulse, such as mode structure, polarization, and pulse shape. In contrast to single-photon sources based on single quantum systems, this attenuation method allows for the use of laser pulses (via mature laser techniques) to produce single photons. Thus we can expect that this method will have many advantages in producing and controlling identical single-photon pulses. 
 
T. Hong thanks Kaoru Shimizu, Fumiaki Morikoshi, Yasuhiro Tokura and K. Inoue for helpful comment and discussions.

\end{document}